\documentclass[twocolumn,showpacs,showkeys,preprintnumbers,amsmath,amssymb]{revtex4}
\usepackage{hyperref}
\usepackage{revsymb}
\usepackage{graphicx}
\usepackage{dcolumn}
\usepackage{bm}
\begin{document}
\draft
\title{Quantum nonlocality for entanglement of quasiclassical states}
\author{Zhi-Rong Zhong, Jian-Qi Sheng, Li-Hua Lin}
\author{Shi-Biao Zheng}
\email{t96034@fzu.edu.cn}
\address{Fujian Key Laboratory of Quantum Information and Quantum Optics,
College of Physics and Information Engineering, Fuzhou University, Fuzhou,
Fujian 350108, China}
\begin{abstract}
Entanglement of quasiclassical (coherent) states of two harmonic oscillators
leads to striking quantum effects and is useful for quantum technologies.
These effects and applications are closely related to nonlocal correlations
inherent in these states, manifested by the violation of Bell inequalities.
With previous frameworks, this violation is limited by the size of the
system, which does not approach the maximum even when the amount of
entanglement approaches its maximum. Here we propose a new version of Bell
correlation operators, with which a nearly maximal violation can be obtained
as long as the associated entanglement approximates to the maximum.
Consequently, the revealed nonlocality is significantly stronger than those
with previous frameworks for a wide range of the system size. We present a
new scheme for realizing the gate necessary for measurement of the nonlocal
correlations. In addition to the use in test of quantum nonlocality, this
gate is useful for quantum information processing with coherent states.
\end{abstract}

\pacs{03.65.Ud, 03.65.Ta, 03.67.Mn}
\vskip 0.5cm
\maketitle

In their seminal paper [1], Einstein, Podolsky, and Rosen concluded that
quantum-mechanical description of reality is not complete based on the
assumption that the properties of an object is not affected by the
measurement on another object that is far away from it, while quantum mechanics
predicts that two or more spatially separated
objects can be in an entangled state so that there exist nonlocal
correlations between these objects. In 1964, Bell constructed an inequality
imposed by local hidden theories, allowing experimental test of quantum
nonlocality by measuring the correlations in the classical properties of
entangled particles [2]. Since then, different versions of Bell inequalities
have been proposed [3-5].

Though the Einstein-Podolsky-Rosen (EPR) paradox was formulated using
entangled states of two systems with infinite-dimensional Hilbert space
(e.g., harmonic oscillators), previous investigations on Bell nonlocality
focused on systems with finite-dimensional space until 1998, when Banaszek
and Wodkiewicz (BW) constructed the Bell inequality for two harmonic
oscillators in terms of the two-mode Wigner function, which is associated
with the joint quantum-number parity of these two systems under a
displacement operation in phase space [6,7]. With this formulation and its
generalization [8], the Bell signal for the original EPR state exceeds
the bound imposed by local hidden theories, but cannot
attain the maximum value of $2\sqrt{2}$ for two-partite entangled states,
known as the Cirel'son bound [9]. Chen et al. [10] formulated a set of
pseudospin operators, which allows correlations in the original EPR state to
maximally violate the Bell inequality; however, measurement of these
pseudospin observables is experimentally challenging.

Among various kinds of entangled states of harmonic oscillators, entangled
coherent states are of particular interest~[11,12]. On one hand, this kind
of states, formed by coherent states with classical analogs, are a two-mode
generalization of Schr\"{o}dinger cat states [13] and can exhibit strong
quantum interference effects, evidenced by fringes and negativity of Wigner
quasiprobability distribution in phase space~{[14-16]}. On the other hand,
they can be considered two-qubit entangled states when the involved coherent
states for each oscillator are approximately orthogonal to each other. In
addition to fundamental interest, these states can be used for quantum
information processing [11] and quantum metrology [17]. In particular, logic
qubits encoded with cat states have inherent error-correctable functions,
offering a promising possibility for implementation of fault-tolerant
quantum computation~[18]. With this encoding, implementation of a quantum
algorithm corresponds to controlled creation and manipulation of
entanglement of coherent states. These nonclassical effects and applications
are closely related to the associated nonlocal correlations. Very recently,
it was shown that quantum nonlocality is responsible for the advantage of
quantum circuits over their classical counterparts [19]. Thus, the
nonlocality of entangled coherent states are both of fundamental interest
and of practical importance. This has been studied with different kinds of
observables~[8,14,20], but with which the
Bell signal approaches $2\sqrt{2}$ only when the size of the system is
large. As the decaying rate of the entanglement scales with the system size,
it is highly desirable to construct a new version of Bell correlation
operators, where the observables are experimentally detectable and with
which stronger nonlocal correlations can be revealed.

In this manuscript, we proposed a new version of Bell correlation operators
for entangled coherent states, with which the uncovered nonlocality is
significantly stronger than those with previous versions for a wide range of
the system size. In our framework, when the coherent states associated with
each oscillator can approximately form a logic qubit, the operators used for
constructing the correlations do not move this qubit out of the logic space.
This is in contrast with previous frameworks, where the cat state qubits
cannot remain in the logic space under the application of the correlation
operators. This distinct feature enables a nearly maximal Bell inequality
violation as long as the entangled coherent state can be approximately
expressed as a two-qubit maximally entangled state. The size of the system
required for approaching the upper bound $2\sqrt{2}$ is reduced by one order
compared with that based on the BW formalism, which is of importance in view
of decoherence. We propose a new scheme for implementing the phase gate
required for measurement of the correlations. Besides fundamental interest,
this gate has applications in quantum information processing with cat state
qubits [21,22].

The entangled coherent states under consideration are defined as
\begin{equation}
\left\vert \psi \right\rangle ={\cal N}\left( \left\vert \alpha
_{1}\right\rangle _{1}\left\vert \alpha _{2}\right\rangle _{2}+e^{i\theta
}\left\vert -\alpha _{1}\right\rangle _{1}\left\vert -\alpha
_{2}\right\rangle _{2}\right) ,
\end{equation}
where ${\cal N}=\left[ 2+2\cos \theta e^{-2\left( \left\vert \alpha
_{1}\right\vert ^{2}+\left\vert \alpha _{2}\right\vert ^{2}\right) }\right]
^{-1/2}$ is the normalization factor, and $\left\vert \alpha
_{j}\right\rangle _{j}=e^{-\left\vert \alpha _{j}\right\vert
^{2}/2}\sum_{n=0}^{\infty }\frac{\alpha _{j}^{n}}{\sqrt{n!}}\left\vert
n\right\rangle _{j}$ denotes the coherent state of amplitude $\alpha _{j}$
for the $j$th ($j=1,2$) harmonic oscillator with $\left\vert n\right\rangle
_{j}$ representing the Fock state with $n$ quanta. The observables we use to
characterize the nonlocal correlations of these entangled states are the
rotated parities, each combined by an effective cat state qubit rotation
operator and the parity operator. The effective cat state qubit rotation
operator associated with the $j$th oscillator is defined as
\begin{equation}
R_{j,z}(\phi _{j})=D_{j}^{\dagger }(\alpha _{j})G_{j}(\phi _{j})D_{j}(\alpha
_{j}),
\end{equation}
where%
\begin{equation}
G_{j}(\phi _{j})=\left\vert 0\right\rangle _{j}\left\langle 0\right\vert
e^{i\phi _{j}}+\sum_{n=1}^{\infty }\left\vert n\right\rangle
_{j}\left\langle n\right\vert
\end{equation}

is the phase gate for the $j$th oscillator that produces a phase shift $\phi
_{j}$ if and only if this oscillator is in the ground state $\left\vert
0\right\rangle _{j}$, and $D_{j}(\alpha _{j})=e^{\alpha _{j}a_{j}^{\dagger
}-\alpha _{j}^{\ast }a_{j}}$ denotes the displacement operator, with $%
a_{j}^{\dagger }$ and $a_{j}$ being the quantum-number rising and lowering
operators. The operation $D_{j}(\alpha _{j})$ translates the quantum state
of the $j$th oscillator by an amount of $\alpha _{j}$ in phase space. The
sequence of operations $D_{j}(\alpha _{j})$, $G_{j}(\phi _{j})$, and $%
D_{j}^{\dagger }(\alpha _{j})$ displaces $\left\vert -\alpha
_{j}\right\rangle _{j}$ to the vacuum state $\left\vert 0\right\rangle _{j}$%
, produces a phase shift $\phi _{j}$ on $\left\vert 0\right\rangle _{j}$,
and finally transforms $\left\vert 0\right\rangle _{j}$ back to $\left\vert
-\alpha _{j}\right\rangle _{j}$. On the other hand, $D_{j}(\alpha _{j})$
evolves $\left\vert \alpha _{j}\right\rangle _{j}$ to $\left\vert 2\alpha
_{j}\right\rangle _{j}$, which is not affected by $G_{j}(\phi _{j})$ and
thus transformed back to $\left\vert \alpha _{j}\right\rangle _{j}$ by $%
D_{j}^{\dagger }(\alpha _{j})$ when $\left\vert \left\langle 0\right\vert
\left. 2\alpha _{j}\right\rangle \right\vert ^{2}\simeq 0$. Therefore, with
the encoding $\left\{ \left\vert 0_{L}\right\rangle _{j}=\left\vert \alpha
_{j}\right\rangle _{j},\left\vert 1_{L}\right\rangle _{j}=\left\vert -\alpha
_{j}\right\rangle _{j}\right\} $, $R_{j,z}(\phi _{j})$ is equivalent to a
rotation of the cat state qubit around the z axis by an angle $\phi _{j}$.
On the other hand, the parity operator $
P_{j}=(-1)^{a_{j}^{\dagger }a_{j}}$ corresponds to the Pauli spin operator $
\sigma _{j,x}$. As a result, the rotated parity operator
\begin{equation}
P_{j,z}(\phi _{j})=R_{j,z}(\phi _{j})P_{j}R_{j,z}^{\dagger }(\phi _{j})
\end{equation}
is analogous to the spin operator along the axis with an angle $\phi _{j}$
to the x axis on the xy plane, denoted as $\sigma _{j,\phi _{j}}$. We note
that such an operator, like the pseudospin operators defined in Ref. [10],
maps the infinite-dimensional Hilbert space onto a two-dimensional space [23].

To construct the Bell inequality, we define the rotated parity-parity
correlation as
\begin{equation}
E(\phi _{1},\phi _{2})=\left\langle P_{1,z}(\phi _{1})\otimes P_{2,z}(\phi
_{2})\right\rangle .
\end{equation}
$E(\phi _{1},\phi _{2})$ corresponds to the mean value of the product of the
spin-like observables of the two cat state qubits, $\sigma _{1,\phi _{1}}$
and $\sigma _{2,\phi _{2}}$. The Bell inequality, based on correlations
defined in this way, is
\begin{equation}
{\cal S}_{RP}=\left\vert E(\phi _{1},\phi _{2})+E(\phi _{1},\phi
_{2}^{^{\prime }})+E(\phi _{1}^{^{\prime }},\phi _{2})-E(\phi _{1}^{^{\prime
}},\phi _{2}^{^{\prime }})\right\vert \leq 2.
\end{equation}%
For the entangled coherent state of Eq. (1), we have

\begin{eqnarray}
\setlength{\abovedisplayskip}{3pt}
\setlength{\belowdisplayskip}{3pt}
E(\phi _{1},\phi _{2}) &=&{\cal N}^{2}\left\{ e^{-2\left( \left\vert \alpha
_{1}\right\vert ^{2}+\left\vert \alpha _{2}\right\vert ^{2}\right)
}+K_{\alpha _{1},\alpha _{1}}(\phi _{1})K_{\alpha _{2},\alpha _{2}}(\phi
_{2})\right.   \nonumber \\
&&\left. +\left[ e^{i\theta }K_{\alpha _{1},-\alpha _{1}}(\phi
_{1})K_{\alpha _{2},-\alpha _{2}}(\phi _{2})+c.c.\right] \right\} ,
\end{eqnarray}
where
\begin{eqnarray}
K_{\alpha _{j},\alpha _{j}}(\phi _{j}) &=&e^{-2\left\vert \alpha
_{j}\right\vert ^{2}}\left( 2\cos \phi _{1}-1\right) +2e^{-6\left\vert
\alpha _{j}\right\vert ^{2}}(1-\cos \phi _{1})  \nonumber \\
K_{\alpha _{j},-\alpha _{j}}(\phi _{j}) &=&e^{-i\phi _{j}}+e^{-4\left\vert
\alpha _{j}\right\vert ^{2}}\left( 1-e^{-i\phi _{j}}\right) .
\end{eqnarray}
The first two terms of $E(\phi _{1},\phi _{2})$ are respectively contributed
by the components $\left\vert -\alpha _{1}\right\rangle _{1}\left\vert
-\alpha _{2}\right\rangle _{2}$ and $\left\vert \alpha _{1}\right\rangle
_{1}\left\vert \alpha _{2}\right\rangle _{2}$, while the last two terms
arise from their quantum coherence, which is responsible for the
entanglement between the two harmonic oscillators. The values of $\left\vert
\alpha _{1}\right\vert $ and $\left\vert \alpha _{2}\right\vert $ determine
the accuracy of the effective cat state qubit rotations. With the increase
of $\left\vert \alpha _{j}\right\vert $,
the rotated parity operator $P_{j,z}(\phi _{j})$ is closer to the spin
operator $\sigma _{j,\phi _{j}}$. When $\left\vert \alpha _{j}\right\vert
\rightarrow \infty $ ($j=1,2$), $E(\phi _{1},\phi _{2})=\cos (\theta
-\phi _{1}-\phi _{2})$, which is in perfect analogy with the correlation
between two photons in a maximally polarization-entangled states [24]. In
this limit, the optimized Bell signal reaches the maximum $2\sqrt{2}$.
 As will be shown, a nearly maximal Bell
violation can be obtained even with quite moderate values of $\left\vert
\alpha _{1}\right\vert $ and $\left\vert \alpha _{2}\right\vert $.
We note that the Bell violation discussed here arises from the quantum
entanglement between two distinct bosonic modes, which is in stark contrast
with situation investigated in a recent experiment [25], where the Bell
violation results from the classical correlation between the polarization
and amplitude of the same optical field.

\begin{figure}
\includegraphics[width=3.5in]{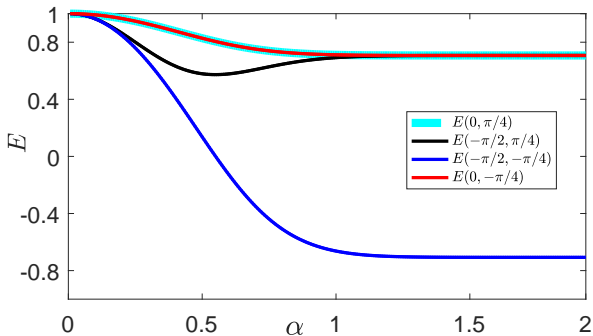}
\caption{(Color online) Numerical simulations of correlations for the 3-mode
cat state as functions of $\alpha $: (a), $E_{3}(0,-\pi /4,\pi /4)$; (b), $%
E_{3}(0,\pi /4,-\pi /4)$; (c), $E_{3}(\pi /2,-\pi /4,-\pi /4)$; (d), $%
E_{3}(\pi /2,\pi /4,\pi /4)$. The solid and dotted lines corrrespond to the
correlations for the cat state and the equally-weighted classical mixture of
the two components $\left\vert \alpha \right\rangle _{1}\left\vert \alpha
\right\rangle _{2}\left\vert \alpha \right\rangle _{3}$ and $\left\vert
-\alpha \right\rangle _{1}\left\vert -\alpha \right\rangle _{2}\left\vert
-\alpha \right\rangle _{3}$, respectively.}
\end{figure}

To show the degree of the Bell inequality violation for limited values of $%
\alpha _{1}$ and $\alpha _{2}$, we perform numerical simulations of the Bell
signal ${\cal S}_{RP}$. For simplicity, we here set $\theta =0$. For this
setting, the choice $\phi _{1}=0$, $\phi _{2}=\pi /4$, $\phi _{1}^{^{\prime
}}=-\pi /2$, and $\phi _{2}^{^{\prime }}=-\pi /4$ ensures ${\cal S}_{RP}$ to
reach the maximum when $\left\vert \alpha _{j}\right\vert \rightarrow \infty
$. As shown in Eqs. (7) and (8), the correlation $E(\phi _{1},\phi _{2})$
has nothing to do with the phases of $\alpha _{1}$ and $\alpha _{2}$, which
will be taken to be positive real numbers in our
simulations. To clearly show the difference between the Bell violation with
the present correlation operators and that with the BW formalism, we first
take $\alpha _{1}=\alpha _{2}=\alpha $. The four correlations $E(0,\pi /4)$,
$E(0,-\pi /4)$, $E(-\pi /2,\pi /4)$, and $E(-\pi /2,-\pi /4)$ as functions
of the value of $\alpha $, are displayed in Fig. 1. As expected, each of the
these correlations is 1 when $\alpha =0$. This is due to the fact that for
this case $\left\vert \psi \right\rangle =\left\vert 0\right\rangle
_{1}\left\vert 0\right\rangle _{2}$ and $P_{j,z}(\phi _{j})\left\vert
0\right\rangle _{j}=\left\vert 0\right\rangle _{j}$, so that $\left\langle
P_{1,z}(\phi _{1})\otimes P_{2,z}(\phi _{2})\right\rangle =1$. With the
increases of $\alpha $, the parity of each oscillator drops quickly with the
other being traced out and the entanglement contributes more and more to the
correlations; $E(-\pi /2,-\pi /4)$ quickly approaches $-\sqrt{2}/2$, while
the other correlations tend to $\sqrt{2}/2$.

The solid line of Fig. 2(a) shows the Bell signal ${\cal S}_{RP}$ calculated
with the rotated parity operators as a function of $\alpha $. The dashed
line represents the maximized Bell signal obtained by the generalized BW
formalism [14], defined as
\begin{equation}
\setlength{\abovedisplayskip}{3pt}
\setlength{\belowdisplayskip}{3pt}
{\cal S}_{BW}=\left\vert E(\beta _{1},\beta _{2})+E(\beta _{1},\beta
_{2}^{^{\prime }})+E(\beta _{1}^{^{\prime }},\beta _{2})-E(\beta
_{1}^{^{\prime }},\beta _{2}^{^{\prime }})\right\vert ,
\end{equation}
where
\begin{equation}
\setlength{\abovedisplayskip}{3pt}
\setlength{\belowdisplayskip}{3pt}
E(\beta _{1},\beta _{2})=\left\langle D_{1}(\beta _{1})P_{1}D_{1}^{\dagger
}(\beta _{1})\otimes D_{2}(\beta _{2})P_{2}D_{2}^{\dagger }(\beta
)\right\rangle
\end{equation}

\begin{figure}
\centering
\includegraphics[width=3.5in]{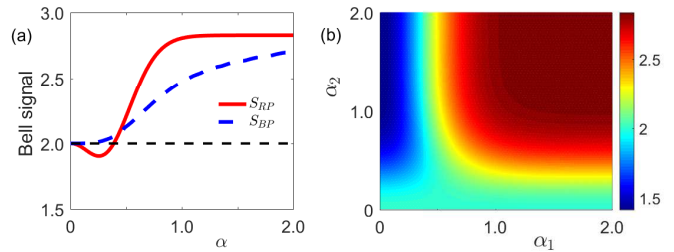}
\vspace{-0.2cm}
\caption{(Color online) MK signal for the 3-mode cat state as a function $%
\alpha $. The solid line represents the numerical result calculated by Eq.
(12), and the dotted line denotes the maximum allowed by local realism. }
\end{figure}

is the expectation value of the joint displaced parity operator of the
two oscillators. The results show that,
when the value of $\alpha $ is smaller than 0.32, the Bell signal obtained
by our formalism decreases as $\alpha $ increases. This can be interpreted
as follows. In this formalism, for a certain value of $\alpha $, all the
four correlations are fixed. In the range $0\leq \alpha <0.32$, the
correlations $E(-\pi /2,\pi /4)$ and $E(-\pi /2,-\pi /4)$ are almost equal
and cancel out each other in the Bell signal, so that ${\cal S}_{RP}$
approximates to the sum of $E(0,\pi /4)$ and $E(0,-\pi /4)$, which drop as $%
\alpha $ increases. On the other hand, in the BW formalism one can choose a
suitable set of displacement parameters for each value of $\alpha $, so that
$E(\beta _{1}^{^{\prime }},\beta _{2}^{^{\prime }})$ drops faster than the
sum of the other three correlations, and consequently ${\cal S}_{BW}$
increases slowly with $\alpha $.

When $\alpha $ exceeds 0.32, the discrepancy
between the correlations $E(-\pi /2,-\pi /4)$ and $E(-\pi /2,-\pi /4)$
enlarges quickly, so that the contribution of their difference to ${\cal S}%
_{RP}$ improves fast. For $\alpha >0.47$, ${\cal S}_{RP}$ surpasses 2. When $%
\alpha >0.55$, the Bell inequality violation obtained by our formalism is
stronger than that by the BW formalism, and ${\cal S}_{RP}$ quickly
approaches the bound $2\sqrt{2}$. The qualitative
interpretation is as follows. Under the application of $R_{j,z}(\phi )$,
logic state $\left\vert 1_{L}\right\rangle _{j}$ undergoes a phase shift $%
\phi _{j}$, while $\left\vert 0_{L}\right\rangle _{j}$ evolves to $%
\left\vert 0_{L}\right\rangle _{j}+e^{-2\left\vert \alpha \right\vert
^{2}}\left( e^{i\phi _{j}}-1\right) \left\vert 1_{L}\right\rangle _{j}$,
whose overlap with $\left\vert 0_{L}\right\rangle _{j}$ approximates unity
even for a moderate value of $\alpha $. For example, this overlap is above
0.999 for $\alpha =\sqrt{2}$. As a result, the operation $R_{j,z}(\phi _{j})$
acts as a nearly perfect rotation around the z axis for the cat state qubit,
which remains in the logic space $\left\{ \left\vert 0_{L}\right\rangle _{j}%
\text{, }\left\vert 1_{L}\right\rangle _{j}\right\} $ after this operation.
On the other hand, in the BW formalism the qubit rotations are replaced by
displacement operations, which move the corresponding harmonics oscillators
out of the logic space for a limited value of $\alpha $ [8]. Consequently,
the value of $\alpha $ required for nearly maximal Bell violation is much
larger than that with our formalism. For example, with the BW formalism the
optimized Bell signal is ${\cal S}_{BW}=2.77$ for $\alpha =3$; based on our
formalism, the Bell signal of the same value can be obtained for $\alpha
\simeq 1$, corresponding to a two-mode cat state with the size reduced by
one order [26]. Since the quantum coherence of a cat
state decays at a rate increasing with its size [26], our
formalism is important for experimental investigation of the quantum
nonlocality for entanglement of quasiclassical states. We note that the
pseudospin flip operators proposed in Ref. [10] also move the cat state
qubits out of the logic space when $\left\vert \alpha \right\vert $ is not
large [8].  Another important feature of our formalism is that the
obtained Bell signal $S_{RP}$  is close to the bound  $2\sqrt{2}$
when the concurrence  $C$ , characterizing the parity-parity
entanglement, approaches its maximum $1$. For example, with the choice
 $\alpha =1.1$  and  $\theta =0$, $C =\frac{
1-e^{-4\left\vert \alpha \right\vert ^{2}}}{1+e^{-4\left\vert \alpha
\right\vert ^{2}}}\simeq 0.984$, and  $S_{RP}\simeq 2.799 $,
while the Bell signal obtained with the BW formalism is only $2.589$.

Further simulations confirm that for $\alpha _{1}\neq \alpha _{2}$ the Bell
signal can also approach $2\sqrt{2}$ with moderate values of $\alpha _{1}$
and $\alpha _{2}$, as shown in Fig. 2(b). For example, when $\alpha _{1}=1.1$
and $\alpha _{2}=1.3$, ${\cal S}_{RP}=2.812$. We note that, for $\theta \neq 0$,
the rotation angles ($\phi _{1}$, $\phi _{2}$, $\phi _{1}^{^{\prime }}$, $%
\phi _{2}^{^{\prime }}$) need to be adjusted so that $C(\phi _{1},\phi
_{2})=C(\phi _{1},\phi _{2}^{\prime })=C(\phi _{1}^{\prime },\phi
_{2})=-C(\phi _{1}^{\prime },\phi _{2}^{\prime })=\pm \sqrt{2}/2$, where $%
C(\phi _{1},\phi _{2})=\cos (\theta -\phi _{1}-\phi _{2})$. With this
adaption, for different values of $\theta $ the Bell signal, as a function
of $\alpha _{1}$ and $\alpha _{2}$, behaves similarly, approximating to $2%
\sqrt{2}$ when $e^{-2\left( \left\vert \alpha _{j}\right\vert
^{2}+\left\vert \alpha _{k}\right\vert ^{2}\right) }\ll 1$ ($j,k=1,2$).

The phase gate $G_{j}(\phi _{j})$ required to detect the Bell signal can be
realized using a qubit dispersively coupled to
the $j$th oscillator and driven by two subsequent pulses, each producing a $%
\pi $ rotation on the qubit conditional on the oscillator state $\left\vert
0\right\rangle _{j}$~[21,22]. With suitable choice of the axes of the two
rotations, the phase gate $G_{j}(\phi _{j})$ with a desired phase shift can
be obtained, as has been experimentally demonstrated in circuit quantum
electrodynamics architectures [21,22]. We here show that such gates can be
implemented with a single square-shaped pulse. In the interaction picture,
the Hamiltonian for such a system is
\begin{equation}
\setlength{\abovedisplayskip}{3pt}
\setlength{\belowdisplayskip}{3pt}
H_{I}=\hbar \chi a_{j}^{\dagger }a_{j}\left\vert e\right\rangle
_{j}\left\langle e\right\vert +\hbar \left[ \Omega _{j}e^{i\delta
_{j}t}\left\vert e\right\rangle _{j}\left\langle g\right\vert +h.c.\right] ,
\end{equation}%
where $\left\vert e\right\rangle _{j}$ and $\left\vert g\right\rangle _{j}$
denote the excited and ground states of the qubit, $\chi $ is the qubit
frequency shift induced by one quantum of the oscillator due to the
dispersive coupling, $\delta _{j}$ is the detuning between the qubit and the
drive with Rabi frequency $\Omega _{j}$, and $h.c.$ denotes the Hermitian
conjugate. The qubit is initially in the ground state $\left\vert
g\right\rangle _{j}$. When the oscillator is initially in the vacuum state $
\left\vert 0\right\rangle _{j}$, the system evolves as
\begin{eqnarray}
\setlength{\abovedisplayskip}{3pt}
\setlength{\belowdisplayskip}{3pt}
\left\vert \psi _{0}(\tau )\right\rangle &&=e^{-i\delta _{j}\tau /2}\left[
\cos \left( \varepsilon _{j}\tau \right) +i\frac{\delta _{j}}{2\varepsilon
_{j}}\sin \left( \varepsilon _{j}\tau \right) \right] \left\vert
g\right\rangle _{j}\left\vert 0\right\rangle _{j}\nonumber\\&&-i\frac{\Omega _{j}}{
\varepsilon _{j}}\sin \left( \varepsilon _{j}\tau \right) e^{i\delta
_{j}\tau /2}\left\vert e\right\rangle _{j}\left\vert 0\right\rangle _{j},
\end{eqnarray}
where $\varepsilon _{j}=\sqrt{\Omega _{j}^{2}+\delta _{j}^{2}/4}$, $\tau $
is the duration of the pulse. With the choice $\delta _{j}=2\Omega _{j}\frac{
\pi -\phi _{j}}{\sqrt{\phi _{j}(2\pi -\phi _{j})}}$ ($\phi _{j}>0$) or $
\delta _{j}=-2\Omega _{j}\frac{\pi +\phi _{j}}{\sqrt{-\phi _{j}(2\pi +\phi
_{j})}}$ ($\phi _{j}<0$) and $\varepsilon _{j}\tau =\pi $, $\left\vert \psi
_{0}(\tau )\right\rangle =e^{i\phi _{j}}\left\vert g\right\rangle
_{j}\left\vert 0\right\rangle _{j}$.

Under the condition $\delta _{j}$, $\Omega _{j}\ll \left\vert \chi
\right\vert $, the qubit almost remains in the state $\left\vert
g\right\rangle _{j}$ throughout the pulse if the oscillator is in a Fock
state $\left\vert n\right\rangle _{j}$ ($n\neq 0$) due to large detunings [27].
As a consequence, the oscillator undergoes a phase
shift $\phi _{j}$ if and only if it is in the vacuum state, which
corresponds to the phase gate $G_{j}(\phi _{j})$. To verify the validity of
the approximation, we perform a numerical simulation on the fidelity of the
system with the initial state $\left\vert \psi _{2\alpha
_{j}}(0)\right\rangle =\left\vert g\right\rangle _{j}\left\vert 2\alpha
_{j}\right\rangle _{j}$ under the Hamiltonian $H_{I}$. The parameters are
set to be $\Omega _{j}=0.2\chi $, $\phi _{j}=\pi /4$, and $\alpha _{j}=2$,
with $\chi >0$. The fidelity, defined as $F=\left\vert \left\langle \psi
_{2\alpha _{j}}(0)\right\vert \left. \psi _{2\alpha _{j}}(t)\right\rangle
\right\vert ^{2}$, as a function of time is presented in Fig. 3, where $%
\left\vert \psi _{2\alpha _{j}}(t)\right\rangle $ represents the state of
the system at time $t$ under the Hamiltonian $H_{I}$. As expected, the
result shows that the state $\left\vert \psi _{2\alpha _{j}}(0)\right\rangle
$ is almost unaffected by the drive; at the time $\tau $ the fidelity is
about 0.998, confirming the high accuracy of the approximation. We note that
the qubit has a small probability of being driven to $\left\vert
e\right\rangle _{j}$ during the pulse, whose population evolution exhibits
collapses and revivals with a period of about $2\pi /\chi $, which account
for the oscillation features of $F$. Since for $n\neq 0$ the drive induces a
slow phase evolution of about $\Omega _{j}^{2}t/(n\chi )$, the curve shows a
slowly descending trend.

\begin{figure}
\centering
\includegraphics[width=3.5in]{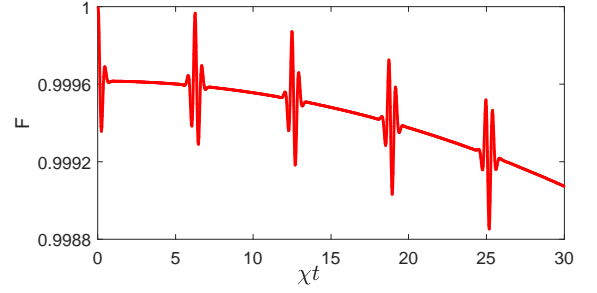}
\caption{(Color online) Numerical simulation of the fidelity evolution with
the initial state $\left\vert g\right\rangle _{j}\left\vert 2\alpha
_{j}\right\rangle _{j}$ under the Hamiltonian of Eq. (11). The parameters
are $\Omega _{j}=0.2\chi $, $\phi _{j}=\pi /4$, and $\alpha _{j}=2$.}
\end{figure}

In conclusion, we have constructed a new version of Bell correlations for
two harmonic oscillators in a two-mode cat state by introducing the
effective cat state qubit rotation operator, which does not move the
corresponding cat state qubit out of the logic space encoded with
quasiorthogonal coherent states. Even when each of the coherent state
components only has a moderate amplitude, the combination of such a rotation
and the quantum number parity measurement is a nearly perfect analog of the
measurement of the spin of a spin-1/2 particle along one prechosen direction
on the xy plane. As a consequence, the Bell signal for a cat state of a
moderate size can approximately attain that for two spin-1/2 particles in a
maximally entangled state, and the revealed nonlocal correlation between the
parities of the two harmonic oscillators coincides with the entanglement in
the mesoscopic regime. We note that such parity-parity entanglement may be
used for quantum communication protocols, e.g., entanglement-based quantum
cryptography [28], with the Bell signal serving as the
measure to test for eavesdropping. We further present a new scheme for
implementation of the phase gate necessary for realizing the cat state qubit
rotation with a qubit dispersively coupled to the corresponding oscillator.
In addition to nonlocality test, this gate has applications in quantum
information processing with coherent states.

This work was supported by the National Natural Science Foundation of China
under Grant No. 11874114 and No. 11674060, the Natural Science Foundation of
Fujian Province under Grant No. 2016J01018, and the Fujian Provincial
Department of Education under Grant No. JZ160422.

\end{document}